\documentstyle[aps,prl,multicol]{revtex}
\begin{document}
\draft

\title{Selective Transport and Mobility Edges \\ in Quasi-1D Systems
       with a Stratified Correlated Disorder}

\author{F.~M.~Izrailev}
\address{Instituto de F\'{\i}sica, Universidad Aut\'{o}noma de Puebla, \\
         Apartado Postal J-48, Puebla, Pue., 72570, M\'{e}xico}

\author{N.~M.~Makarov}

\address{Instituto de Ciencias, Universidad Aut\'{o}noma
         de Puebla, \\ Priv. 17 Norte No 3417, Col. San Miguel
         Hueyotlipan, Puebla, Pue., 72050, M\'{e}xico}

\date{\today}
\maketitle

\begin{abstract}
We present analytical results on transport properties of many-mode
waveguides with randomly stratified disorder having long-range
correlations. To describe such systems, the theory of 1D transport
recently developed for a correlated disorder is generalized. The
propagation of waves through such waveguides may reveal a quite
unexpected phenomena of a complete transparency for a subset of
propagating modes. We found that with a proper choice of
long-range correlations one can arrange a perfect transparency of
waveguides inside a given frequency window of incoming waves.
Thus, mobility edges are shown to be possible in quasi-1D geometry
with correlated disorder. The results may be important for
experimental realizations of a selective transport in application
to both waveguides and electron/optic nanodevices.
\end{abstract}


\begin{multicols}{2}

During last few years there is a burst of interest to the problem
of localization-delocalization transition in systems with
correlated disorder (see, e.g., \cite{nature,our} and references
therein). This fact is due to the possibility to observe an
anomalous transport in 1D models with random potentials. In
particular, it was shown \cite{our} that specific long-range
correlations give rise to a complete transparency of electron
waves for given energy intervals. Experimental realization of such
potentials for single-mode waveguides with delta-like scatters
\cite{KIKS00} has confirmed theoretical predictions.

The subject of wave propagation through disordered many-mode
waveguides is important both from the theoretical viewpoint and
for experimental applications such as optic fibers, remote
sensing, radio wave propagation, shallow water waves, etc (see, for
example, \cite{applic}). It has also a direct relation to the
problem of electronic transport in mesoscopic conducting channels.
So far, main results in this field have been obtained for random
potentials of white-noise type. However, there are many physical
situations where this assumption is not correct. Therefore, the
understanding of generic properties of transport in the models
with correlated disorder is important for the modern theory of
electron/wave propagation. It should be also stressed that
existing experimental technics allow for the construction of
systems with sophisticated scattering potentials resulting in
anomalous transport properties \cite{exper}. 

In this paper we study transport properties of quasi-1D waveguides
with a {\it stratified} bulk disorder that has long-range correlations.
Our main interest is in exploring the possibility of constructing
such random potentials that result in frequency windows of a
perfect propagation of waves.

In what follows we consider a plane waveguide (or conducting wire)
of width $d$, stretched along the $x$-axis. The $z$-axis is
directed perpendicularly to the waveguide so that its lower edge
is $z=0$ and the upper one is $z=d$. The waveguide is assumed to
have a stratified (along the $x$-axis) region of the length $L$
($|x|<L/2$). As a physically plausible model for such a disorder
we consider random potential $V(x)$ with the zero average,
$\langle V(x)\rangle=0$. The angular brackets here and below stand
for the statistical average over realizations of the random
function $V(x)$. Its correlator $\langle V(x)V(x')\rangle ={\cal
W}(|x-x'|)$ is assumed to have a maximal value at $x=x'$ and to
decrease with an increase of $|x-x'|$. For statistical treatment
of the problem it is also important to assume that the
characteristic scale of the decrease of ${\cal W}(|x-x'|)$ is much
less than the length $L$ of the scattering region.

We shall characterize transport properties of the waveguide by its
{\it transmittance} $T(L)$ that within the linear response theory
is defined by the Kubo's formula \cite{Kubo57},
\begin{equation}\label{T(L)-def}
T(L)=-\frac{4}{L^2}\int d\vec{r}\,d\vec{r'}\,\,
\frac{\partial{\cal G}(\vec{r},\vec{r'})}{\partial x}
\frac{\partial{\cal G}^*(\vec{r},\vec{r'})}{\partial x'}.
\end{equation}
Here the integration with respect to $\vec{r}=(x,z)$ runs over the
disordered region. Since the potential $V(x)$ does not depend on
$z$, the retarded Green function has the form,
\begin{equation}\label{GF}
{\cal G}(\vec{r},\vec{r'})=\frac{2}{d}\sum_{n=1}^{N_d}
\sin\left(\frac{\pi nz}{d}\right)
\sin\left(\frac{\pi nz'}{d}\right)G_{n}(x,x')\,.
\end{equation}
Here $N_d=[kd/\pi]$ is the total number of propagating waveguide
modes ({\it conducting channels}) determined by the integer part
$[...]$ of the ratio $kd/\pi$. The wave number $k$ is equal to
$\omega/c$ for a classical scalar wave of frequency $\omega$, and
to the Fermi wave number for electrons.

By substituting Eq.~(\ref{GF}) into Eq.~(\ref{T(L)-def}), we come
to the Landauer's formula \cite{Land92},
\begin{equation}\label{T-tot}
T(L)=\sum_{n=1}^{N_d}T_n(L),
\end{equation}
with $T_n(L)$ being obtained from Eq.~(\ref{T(L)-def}) in which
the integration is performed over $x$, $x'$ only, and
${\cal G}$ is replaced by $G_{n}$. The mode Green
function $G_{n}(x,x')$ is entirely determined by the 1D equation
with the random potential $V(x)$ and lengthwise wave number
$k_n=\sqrt{k^2-(\pi n/d)^2}$,
\begin{equation}\label{GFn-eq}
\left[\frac{d^2}{dx^2}+k_n^2-V(x)\right]\,G_{n}(x,x')=\delta(x-x').
\end{equation}

From Eq.~(\ref{T-tot}) one can see that the {\it total
transmittance} $T(L)$ of a quasi-1D stratified structure is
expressed as a sum of {\it partial transmittances} $T_n$. Every
transmittance $T_n$ describes the transparency of the
corresponding $n$th propagating mode. In such a way we reduce the
transport problem for the quasi-1D disordered system to the 1D
random model (\ref{GFn-eq}) of wave scattering in every $n$th
channel. Eq.~(\ref{GFn-eq}) can be solved by one of the methods
developed in the transport theory of one-dimensional disordered
systems, such as the perturbative diagrammatic method of
Berezinskii \cite{Ber73}, invariant imbedding method
\cite{BelWing75}, or the two-scale approach \cite{MT98}.

The main result is that the average partial transmittance $\langle
T_n\rangle$ as well as all its moments, are described by the
universal function that depends on one parameter $\Lambda_n\equiv
L/L_{loc}(k_n)$ only ({\it one-parameter scaling}). The quantity
$L_{loc}(k_n)/4$ is, in fact, the backscattering length for $n$-th
propagating mode. The inverse value $L_{loc}^{-1}(k_n)$ is equal
to the corresponding Lyapunov exponent that can be found in the
transfer matrix approach, so that the quantity $L_{loc}(k_n)$ is
the {\it localization length} \cite{LGP88} associated with a given
1D channel.

Since general expression for the mode transmittance has quite
complicated form (see, e.g., Ref.~\cite{MT98}), here we refer to
limit cases only. Specifically, for large localization length,
$\Lambda_n=L/L_{loc}(k_n)\ll 1$, the transmittance $\langle
T_n\rangle$ exhibits the ballistic behavior and the corresponding
$n$th normal mode is practically transparent,
\begin{equation}\label{Tn-bal}
\langle T_n\rangle\approx 1-4\Lambda_n.
\end{equation}
On the contrary, the transmittance is exponentially small when the
localization length is much less than the length of the waveguide,
$\Lambda_n\gg 1$,
\begin{equation}\label{Tn-loc}
\langle T_n\rangle\approx \frac{\pi^{5/2}}{16}\Lambda_n^{-3/2}
\exp (-\Lambda_n).
\end{equation}
This implies strong wave localization in a given $n$th channel.
Since in this case the transmittance $T_n$ is not a self-averaged
quantity, it is more reasonable to deal with its average
logarithm, $\langle\ln T_n\rangle=-4\Lambda_n$.

Thus, in order to describe transport properties of a quasi-1D
waveguide with 1D bulk disorder, the  value of localization
lengths $L_{loc}(k_n)$ should be known. From the solution of the
equation (\ref{GFn-eq}) one can obtain
\cite{Ber73,BelWing75,MT98},
\begin{equation}\label{Lloc}
L_{loc}^{-1}(k_n)=\frac{W(2k_n)}{16k_n^2},
\end{equation}
where $W(k_x)$ is the Fourier transform of the binary
correlator ${\cal W}(|x-x'|)$ for the scattering potential,
\begin{equation}
{\cal W}(|x-x'|)=\int_{-\infty}^{\infty}\frac{dk_x}{2\pi}
\exp[ik_x(x-x')]\,W(k_x).
\label{FR-W}
\end{equation}

The main feature of the expression (\ref{Lloc}) for $L_{loc}(k_n)$
is its dependence on the mode index $n$. One can see that the
larger $n$ is, the smaller is $L_{loc}(k_n)$ and, consequently,
the stronger is the coherent scattering within this mode. This
dependence is quite strong due to the squared wave number $k_n$ in
the denominator of Eq.~(\ref{Lloc}). Evidently, with an increase
of the index $n$ the value of $k_n$ decreases. An additional
dependence appears because of the power spectrum $W(2k_n)$. Since
the correlator ${\cal W}(|x-x'|)$ is a decreasing function of
$|x-x'|$, the numerator $W(2k_n)$ increases with $n$ (it is a
constant in the case of the delta-correlated potential only).
Therefore, {\it both} the numerator and denominator contribute in
the same way for the dependence of $L_{loc}(k_n)$ on $n$. As a
result, we arrive at the {\it concept of hierarchy of mode
localization lengths},$$L_{loc}(k_{N_d})<L_{loc}(k_{N_d-1})<...
<L_{loc}(k_2)<L_{loc}(k_1).$$

Thus, a remarkable phenomenon arises. On the one hand, the concept
of one-parameter scaling holds for any of $N_d$ channels whose
partial transport is characterized solely by the value of
$\Lambda_n$. On the other hand, this concept turns out to be
broken for the total waveguide transport. Indeed, due to the
revealed hierarchy of $L_{loc}(k_n)$, the {\it total}
transmittance (\ref{T-tot}) depends on the whole set of scaling
parameters $\Lambda_n$. This fact is in contrast with quasi-1D
bulk-disordered models, for which all transport properties are
described by one parameter only. In this sense, our model is
similar to quasi-1D waveguides with rough surfaces, for which the
similar hierarchy of attenuation lengths was recently found
\cite{SMFY99,Garcia98}.

The interplay between the hierarchy of $L_{loc}(k_n)$ and the
one-parameter scaling for $\langle T_n\rangle$ leads to that, in
general, quasi-1D stratified structures exhibit three transport
regimes.

(I) If the largest of the localization lengths, $L_{loc}(k_1)$, is
much less than the waveguide length $L$,
\begin{equation}\label{tot-loc}
L_{loc}(k_1)\ll L,
\end{equation}
all the propagating modes are localized and the waveguide is
non-transparent.

(II) On the contrary, when the smallest localization length
$L_{loc}(k_{N_d})$ is much larger than $L$,
\begin{equation}\label{tot-trans}
L_{loc}(k_{N_d})\gg L,
\end{equation}
all the propagating modes are open ($\langle T_n\rangle\approx 1$)
and the waveguide has almost perfect transparency. The total
ballistic transmittance in this case is equal to the total number
of the propagating modes, $\langle T(L)\rangle=N_d$.

(III) The intermediate situation arises when $L_{loc}(k_1)$ of the
first mode is larger, while $L_{loc}(k_{N_d})$ of the "last"
$N_d$-th mode is smaller than the waveguide length $L$,
\begin{equation}\label{intm}
L_{loc}(k_{N_d})\ll L\ll L_{loc}(k_1).
\end{equation}
In this case a very interesting phenomenon of the {\it
coexistence} of {\it ballistic} and {\it localized} transport
occurs. Namely, while {\it lowest} modes are in the ballistic
regime, {\it highest} modes are strongly localized.

As a demonstration, let us consider waveguides with a large number
of conducting channels, $N_d=[kd/\pi]\approx kd/\pi\gg 1$, and
with potentials having the widely used Gaussian correlator,
\begin{eqnarray}
{\cal W}(|x|)&=&W_0k_0\pi^{-1/2}\exp\left(-k_0^2x^2\right),
\nonumber\\
W(k_x)&=&W_0\exp\left(-k_x^2/4k_0^2\right).
\label{cor-Gaus}
\end{eqnarray}
It is convenient to introduce two parameters
\begin{equation}
\alpha\equiv\frac{L}{L_{loc}^{w}(k_1)}=\frac{W_0L}{16k^2},
\quad\delta\equiv\frac{L_{loc}^{w}(k_{N_d})}{L_{loc}^{w}(k_1)}=
\frac{2\{kd/\pi\}}{(kd/\pi)}\ll 1,
\label{aplha-delta}
\end{equation}
where $L_{loc}^{w}(k_1)$ and $L_{loc}^{w}(k_{N_d})$ refer,
respectively, to the largest and the smallest localization lengths
in the limit case of the white-noise potential ($k_0\to\infty$),
and $\{kd/\pi\}$ is the fractional part of the parameter $kd/\pi$.

By applying the above discussed approach, one can find that for
Gaussian correlations (\ref{cor-Gaus}) all the propagating modes
are localized when $\alpha\gg\exp(k^2/k_0^2)$. This inequality is
stronger than that valid for the white-noise case, $\alpha\gg 1$.
The intermediate situation occurs when
\begin{equation}\label{GC-intm}
\exp(k^2/k_0^2)\gg\alpha\gg\delta\exp(\delta k^2/k_0^2).
\end{equation}
One can see that the longer range $k_0^{-1}$ of the correlated
disorder is, the simpler are the conditions (\ref{GC-intm}) of the
{\it coexistence} of ballistic and localized transport.

Finally, the waveguide is almost perfectly transparent in the case
when $\alpha\ll\delta\exp(\delta k^2/k_0^2)$. In particular, at
any given value of $\alpha\gg 1$ (fixed values of the waveguide
length $L$, wave number $k$ and the disorder strength $W_0$) the
ballistic transport of all the conducting channels can be realized
by a proper choice of the stratified correlated disorder. From
Eq.~(\ref{GC-intm}) one can also understand under what conditions
the coexistence of ballistic and localized modes can be achieved
for the standard white-noise case.

The fundamentally different situation arises when the stratified
medium has specific long-range correlations. To show this, we
would like to note that the localization length $L_{loc}(k_n)$ of
any $n$th conducting channel are entirely determined by the
dependence $W(k_x)$, see Eq.~(\ref{Lloc}). Therefore, if $W(2k_n)$
abruptly vanishes within some interval of wave number $k_n$, then
$L_{loc}(k_n)$ diverges and the corresponding propagating mode
appears to be fully transparent for any length of the waveguide.
Thus, it is naturally to ask how to construct such random profiles
of stratified medium that result in a complete transparency of
waveguides within any desired frequency region. The answer to this
question can be found from a generalization of the methods
developed in Refs.~\cite{our,IzMak02}. Specifically, having a
desirable dependence for the randomness power spectrum $W(k_x)$,
one should obtain the function $\beta(x)$ whose Fourier transform
is $W^{1/2}(k_x)$. Then, the random potential $V(x)$ can be
constructed as a convolution of white noise $Z(x)$ with the
function $\beta(x)$,
\begin{equation}\label{V-beta}
V(x)=\int_{-\infty}^\infty\,dx'\,Z(x-x')\,\beta(x').
\end{equation}
We emphasize that the transition between localized and ballistic
wave/electron transport can be arranged in an abrupt way at any
given point inside the allowed interval for $k_n$. In order to
achieve such a transition, it is convenient to take $W(k_x)$ with
a discontinuity at the desired point. In other words, the random
potential $V(x)$ should be of a {\it specific} form with {\it
long-range correlations} along the waveguide.

To show how to arrange a sharp transition, let us take a random
potential $V(x)$ with the following power spectrum $W(k_x)$,
\begin{equation}\label{FTW-k0}
W(k_x)=W_0\,\Theta(|k_x|-2k_0),
\end{equation}
where $\Theta(x)$ is the unit-step function and the characteristic
wave number $k_0>0$ is a correlation parameter to be specified.
According to Eq.~(\ref{V-beta}), the random potentials having such
correlator can be constructed as
$$V(x)=W_0^{1/2}\Big[Z(x)-\int_{-\infty}^\infty\,dx'\,
Z(x-x')\,\frac{\sin(2k_0x')}{\pi x'}\Big].$$

Now one can see that in the case under consideration all low modes
with wave numbers $k_n>k_0$ have finite localization lengths while
for high modes with $k_n<k_0$ the localization lengths diverge,
\begin{equation}\label{Llock0}
L_{loc}^{-1}(k_n>k_0)=W_0/16k_n^2,\qquad L_{loc}^{-1}(k_n<k_0)=0.
\end{equation}
Remarkably, in contrast to the potentials with Gaussian
correlations (see above), all propagating modes with
\begin{equation}\label{cond-k0}
n>N_{loc}=[(kd/\pi)(1-k_0^2/k^2)^{1/2}]\Theta(k-k_0),
\end{equation}
exhibit the ballistic behavior. Since their mode transmittance
$T_n=1$, such modes form a {\it coset of completely transparent
channels}. As for other propagating modes with {\it low} indices
$n\leq N_{loc}$, they remain to be {\it localized} for large
enough length size $L$ when the condition (\ref{tot-loc}) holds,
i.e. for $W_0L/16k_1^2\gg 1$.

The expression (\ref{cond-k0}) determines the total number
$N_{loc}$ of {\it localized modes} and total number
$N_{tr}=N_d-N_{loc}$ of {\it completely transparent modes}. Since
localized modes do not contribute to the total transmittance
(\ref{T-tot}), the latter is equal to the number $N_{tr}$ of
completely transparent modes and do not depend on the waveguide
length $L$,
\begin{equation}\label{Ttot-k0}
\langle T\rangle=[kd/\pi]-
[(kd/\pi)(1-k_0^2/k^2)^{1/2}]\Theta(k-k_0).
\end{equation}
We remind that square brackets stand for the integer part of the
inner expression.

For $k_0\ll k$ the number of localized modes
$N_{loc}\approx[(kd/\pi)(1-k_0^2/2k^2)]$ is of the order of $N_d$.
Consequently, the number of transparent modes $N_{tr}$ is small,
or there are no such modes at all. Otherwise, if $k_0\to k$, the
integer $N_{loc}\approx[\sqrt{2}(kd/\pi)(1-k_0/k)^{1/2}]$ turns
out to be much less than the total number of waveguide modes
$N_d$, and the number of transparent modes $N_{tr}$ is large. When
$k_0>k_1$, the number $N_{loc}$ vanishes and {\it all modes}
become fully transparent. In this case the correlated disorder
results in a perfect transmission of waves. One can see that the
value $k_0$ is, in essence, the {\it total} mobility edge that
separates the region of complete transparency from that where
lower modes are localized.

In conclusion, we have studied quasi-1D waveguides with a
stratified disorder that has long-range correlations. We have
shown that with a proper choice of correlations one can arrange a
complete transparency of transverse channels in a given frequency
window of incoming waves. It is worthwhile to
emphasize a non-monotonic step-wise dependence of
the total transmittance (\ref{Ttot-k0}) on the wave number $k$. 
Specifically, within the region $k<k_0$ for
fixed values of $k_0$
(when $ W_0Ld^2/16\pi^2\gg(k_0d/\pi)^2\gg 1$), 
all propagating modes are transparent and the transmittance
exhibits step-wise increase with an increase of $k$. Each step up
arises for an integer value of $kd/\pi$ when new conducting channel
emerges. This effect is similar to that known to occur 
in quasi-1D ballistic (non-disordered)
structures (see, e.g., \cite{vWvHB88}). 

On the other hand, for $k > k_0$
the transmittance reveals a step down decrease
due to successive localization of low modes.
In contrast with the steps up, every step down occurs at the {\it
local} mobility edge of the corresponding channel, when the second
term in Eq.~(\ref{Ttot-k0})
takes the integer value. In general,
these values do not coincide with the integer values of $kd/\pi$,
thus resulting in a new kind of step-wise dependence for the
transmittance.
Finally, for very large  values $k^2\gg W_0L/16$, the transmittance
starts to increase again, due to a successive delocalization of the 
modes.
Our results can be used for fabrications of
electron or optic nanodevices with a selective transport.

This research was supported by CONACYT grant 34668-E, and by BUAP
grant II-104G02 (M\'exico).

\end{multicols}

\end{document}